\documentclass[]{interact}

\usepackage{epstopdf}% To incorporate .eps illustrations using PDFLaTeX, etc.
\usepackage[caption=false]{subfig}% Support for small, `sub' figures and tables

\usepackage[numbers,sort&compress,merge]{natbib}% Citation support using natbib.sty
\bibpunct[, ]{[}{]}{,}{n}{,}{,}% Citation support using natbib.sty
\theoremstyle{plain}% Theorem-like structures provided by amsthm.sty

\theoremstyle{definition}

\theoremstyle{remark}

\usepackage[usenames, dvipsnames]{color}
\usepackage[T1]{fontenc}
\usepackage{pdfpages}

\begin{document}

%\articletype{ARTICLE TEMPLATE}% Specify the article type or omit as appropriate

\title{Reaction rate theory for supramolecular kinetics: application to protein aggregation}

\author{
\name{Thomas C.~T. Michaels\textsuperscript{a}\thanks{CONTACT A.S. and T.P.J.K. Email: a.saric@ucl.ac.uk; tpjk2@cam.ac.uk; LXL and SC contributed equally to this work.}, Lucie X. Liu\textsuperscript{a}, Samo Curk\textsuperscript{c}, Peter G. Bolhuis\textsuperscript{d}, An\dj ela~\v{S}ari\'c\textsuperscript{c} and Tuomas P.~J. Knowles\textsuperscript{a,e}}
\affil{\textsuperscript{a}Department of Chemistry, University of Cambridge, Lensfield Road, Cambridge, CB2 1EW, UK; \textsuperscript{b}Paulson School of Engineering and Applied Sciences, Harvard University, Cambridge, MA 02138, USA; \textsuperscript{c}Department of Physics and Astronomy, Institute for the Physics of Living Systems, University College London, Gower Street, London WC1E 6BT, UK; \textsuperscript{d}van't Hoff Institute for Molecular Sciences, University of Amsterdam, PO Box 94157, 1090 GD Amsterdam, The Netherlands; \textsuperscript{e}Cavendish Laboratory, Department of Physics, University of Cambridge, J J Thomson
Avenue, Cambridge CB3 1HE, UK}
}

\maketitle

\begin{abstract}
Probing the reaction mechanisms of supramolecular processes in soft- and biological matter, such as protein aggregation, is inherently challenging. These processes emerge from the simultaneous action of multiple molecular mechanisms, each of which is associated with the rearrangement of a large number of weak bonds, resulting in a complex free energy landscape with many kinetic barriers. Reaction rate measurements of supramolecular processes at different temperatures can offer unprecedented insights into the underlying molecular mechanisms and their thermodynamic properties. However, to be able to interpret such measurements in terms of the underlying microscopic mechanisms, a key challenge is to establish which properties of the complex free energy landscapes are probed by the reaction rate. Here, we present a reaction rate theory for supramolecular kinetics based on Kramers rate theory for diffusive reactions over multiple kinetic barriers, and apply the results to protein aggregation. Using this framework and Monte Carlo simulations, we show that reaction rates for protein aggregation are of the Arrhenius-Eyring type and that the associated activation energies probe only one relevant barrier along the respective free energy landscapes. We apply this advancement to interpret, both in experiments and in coarse-grained computer simulations, reaction rate measurements of amyloid aggregation kinetics in terms of the underlying molecular mechanisms and associated thermodynamic signatures. Our results establish a general platform for probing the mechanisms and energetics of supramolecular phenomena in soft- and biological matter using the framework of chemical kinetics.
\end{abstract}


\section{Introduction and motivation}
The mechanisms of macromolecular reactions in soft and 
biological matter, such as protein protein association or 
protein aggregation, are notoriously difficult to probe in experiments. 
This difficulty originates from the fact that these complex 
macromolecular processes involve the concurrent making and breaking of 
very large numbers of bonds and interactions between the multiple 
molecular species present. Historically, the key for investigating molecular mechanisms of small molecule reactions
has been to probe the underlying free energy landscape by measuring the 
temperature dependence of the reaction rates. Reaction rate theory then provides the framework for relating these measurements to the thermodynamics of the underlying free energy landscape. 

The discipline of rate theory was established when Arrhenius \cite{Arrhenius_1889} 
described the temperature dependence of the rate $k$ of a chemical 
reaction in terms of what is now known as the Arrhenius equation: $k= 
\nu\, e^{-\beta\Delta G^\ddag}$, where $\nu$ is a frequency pre-factor, 
$\beta=1/k_BT$ is the inverse temperature and $\Delta G^\ddag$ is the 
free energy barrier. The next substantial development came with Eyring \cite{Eyring_1935} 
in the 1930's, who assumed that the reaction is governed by a rate determining step which corresponds to the breaking 
of a single quantum mechanical chemical bond. This assumption allows 
explicit calculation of the frequency prefactor as $\nu = k_BT/h$, where 
$h$ is the Planck constant. Eyring's equation has proved very successful 
in describing the reactions of small molecules, but it is not expected 
to apply to supramolecular processes involving  macromolecules, as these processes require the rearrangement of large numbers of bonds rather than
breaking of a single quantum-mechanical mode of vibration (Fig.~1). Hence, the associated energy landscape in this case involves many kinetic barriers along the reaction coordinate. A more physically realistic model for these systems is offered by Kramers 
rate theory \cite{Kramers_1940,Hanngi_1990}. In this theory, reactions are described as diffusion processes along a complex free energy landscape which is 
parameterized by just a few important coordinates. The reaction rate 
corresponds to the inverse of the escape time and it is found that the 
reaction rate is of the Arrhenius-Eyring 
type, whereby the prefactor depends on key features of the free energy 
landscape. In particular, it is found that $\nu=\omega_1\omega_2/(2\pi 
\gamma)$, where $\omega_1$ and $\omega_2$ denote the curvature of the 
potential landscape at the bottom and the top of the free energy 
barrier, respectively, and $\gamma$ is the friction coefficient.

Characterizing the molecular mechanisms of macromolecular diffusive processes thus requires solving the inverse problem of characterising the free energy landscape by measuring the reaction rate. Clearly, the reaction rate will contain the information about the associated free energy barrier, $\Delta G^\ddag/k_BT$, via Kramers equation. Specifically, by measuring the temperature-variation of the reaction rate, the information about the free energy barrier becomes directly accessible. A key question therefore is to establish which free energy barrier on a complex multibarrier landscape is read out by such a measurement. 

Here, we review and apply Kramers reaction theory to model the molecular mechanisms of reactions governed by diffusive dynamics and, conversely, to establish which information about the free energy landscape can be obtained from the temperature-dependence of the associated reaction rate. We find that only one specific barrier from the multi-barrier landscape is probed by such measurements. We then apply this framework to study the energetics of protein aggregation phenomena, a biologically relevant example of multi-step diffusive processes with implications in areas ranging from biomedicine \cite{Chiti_2017} to nanotechnology \cite{Gazit_2007,Knowles_2011,Wei_2017,Mezzenga_2013}. Specifically, we apply Kramer's theory in conjunction with coarse-grained computer simulations and kinetic experiments to determine the thermodynamic characteristics of key steps involved in the aggregation of Alzheimer's Amyloid-$\beta$ peptide into amyloid fibrils. These results establish a general platform for probing the energetics of complex macromolecular reactions in soft matter.

\begin{figure} 
\includegraphics[width=\textwidth]{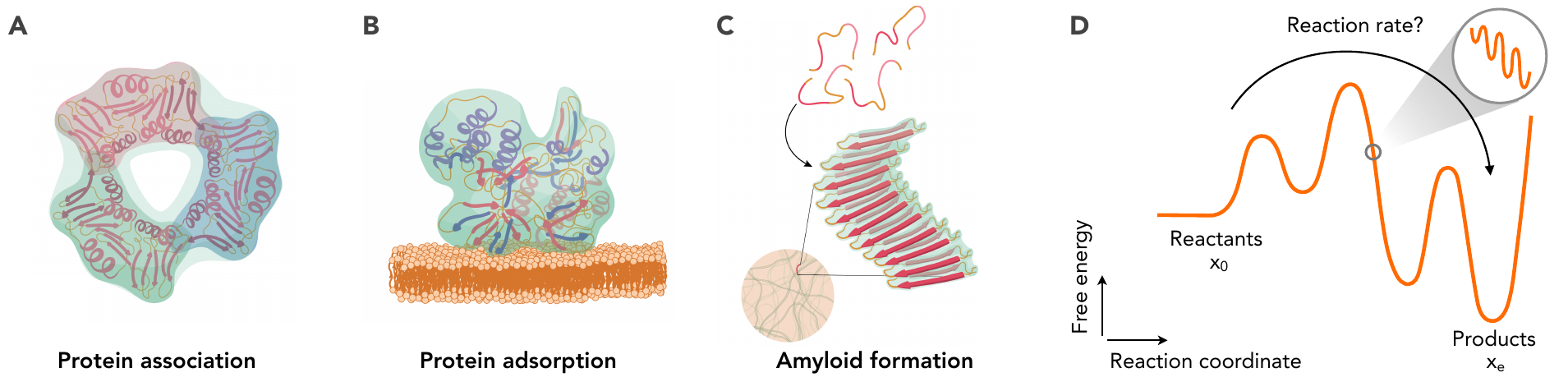}
\caption{\textbf{Examples of supramolecular kinetics in soft and biological matter.} (a) Formation of protein complexes, (b) protein adsorption, (c) amyloid fibril formation. (d) Supramolecular kinetics are characterized by a complex free energy landscape with multiple kinetic barriers. A key question is to establish which features of the detailed free energy landscape are probed by measurements of the overall reaction rate.}
 \end{figure}
 
 \section{Kramers theory of diffusive reactions with multiple kinetic barriers}

%\subsection{Diffusive reaction model for macromolecular processes}
Let us consider a diffusive reaction between well-defined initial and final states, $x_0$ and $x_e$, taking place over a one-dimensional potential free energy landscape, $ G(x)$, with multiple barriers (Fig.~1d). 

 The following Fokker-Planck equation then describes the time evolution of the probability $p(x,t|x_0)$ that, starting at $x_0$, the system has diffused to position $x$ at time $t$ \cite{Hanngi_1990}:  
\begin{equation}\label{FP}
\frac{\partial p(x,t|x_0)}{\partial t} = -\frac{\partial \mathcal{J}}{\partial x}
\end{equation}
where 
\begin{equation}
\mathcal{J} = -\frac{1}{\gamma}\frac{\partial G(x)}{\partial x} p(x,t|x_0)-D \ \frac{\partial p(x,t|x_0)}{\partial x}.
\end{equation}
Here, $\gamma$ denotes the frictional coefficient and $D$ is the diffusion coefficient. Note that $\gamma$ and $D$ are related to the thermal energy through the Einstein-Smoluchowski relation, $\gamma D = k_B T$.

%\subsection{Arrhenius-Eyring behaviour for rate constants of protein aggregation}

The key quantity of interest is the average first passage time $\tau(x_0\to x_e)$, i.e. the time that it takes, on average, for a system described by Eq.~\eqref{FP} and starting at $x_0$ to reach $x_e$. 
In fact, the inverse of the average first passage time corresponds to the transition rate from $x_0$ to $x_e$ \cite{Reimann}

\begin{equation}\label{k}
k(x_0\to x_e)= \frac{1}{\tau(x_0\to x_e)},
\end{equation}
thus providing a link between free energy landscapes and experimental reaction rate measurements. The average first passage time $\tau(x_0\to x_e)$ is related to the probability $p(x,t|x_0)$ through (see Supplementary Material)
\begin{align}\label{tau}
 \tau(x_0\to x_e) = \int_0^\infty  dt \int_{-\infty}^{x_e} dx \ p(x,t|x_0). 
 \end{align}
Integrating the Fokker-Planck equation \eqref{FP} using Eq.~\eqref{tau}, we find that $\tau(x_0\to x_e)$ satisfies the following differential equation (see Supplementary Material):
  \begin{equation}\label{ode_tau}
 -\frac{1}{\gamma} \frac{\partial  G(x)}{\partial x} \ \frac{\partial  \tau(x\to x_e)}{\partial x} + D\ \frac{\partial^2 \tau(x\to x_e)}{\partial x^2} = -1.
 \end{equation}
 The solution to Eq.~\eqref{ode_tau} subject to the boundary condition $\tau(x_e \to x_e) =0$ is (see Supplementary Material):
 \begin{equation}\label{tau_final}
 \tau(x_0\to x_e) = \beta \gamma \int_{x_e}^{x_0} dy \int_{-\infty}^y dz \ e^{\beta[  G(y)-  G(z)]}.
 \end{equation}
In the limit when the relevant free energy barrier is much bigger than thermal energy ($ \beta  \Delta G \gg 1$), the integrals in Eq.~\eqref{tau_final} can be evaluated explicitly using the saddle point approximation \cite{Bender}. In particular, assuming that the width does not vary significantly between the multiple potential energy barriers, we need to maximize the integrand over the integration range of Eq.~\eqref{tau_final}, i.e. we need to find $\max_{z\leq y}[  G(y)-  G(z)]$. Let $y=x^*$ and $z=x_*$ denote the points in the range of integration of Eq.~\eqref{tau_final} where the integrand in Eq.~\eqref{tau_final} is maximal. We find (see Supplementary Material): 
\begin{equation}\label{t}
\tau(x_0\to x_e)  \simeq \frac{2\pi \gamma}{\omega_1\omega_2}  e^{\beta  \Delta G^\ddag},
\end{equation}
where
\begin{equation}\label{rate2}
 \Delta G^\ddag = \max_{x_0\leq z\leq y \leq x_e}[  G(y) - G(z)]
\end{equation}
and $\omega_1$ and $\omega_2$ are the curvatures of the free energy landscape at $x_*$ and $x^*$, respectively.

\begin{figure}
\begin{center}\includegraphics[width=0.6\textwidth]{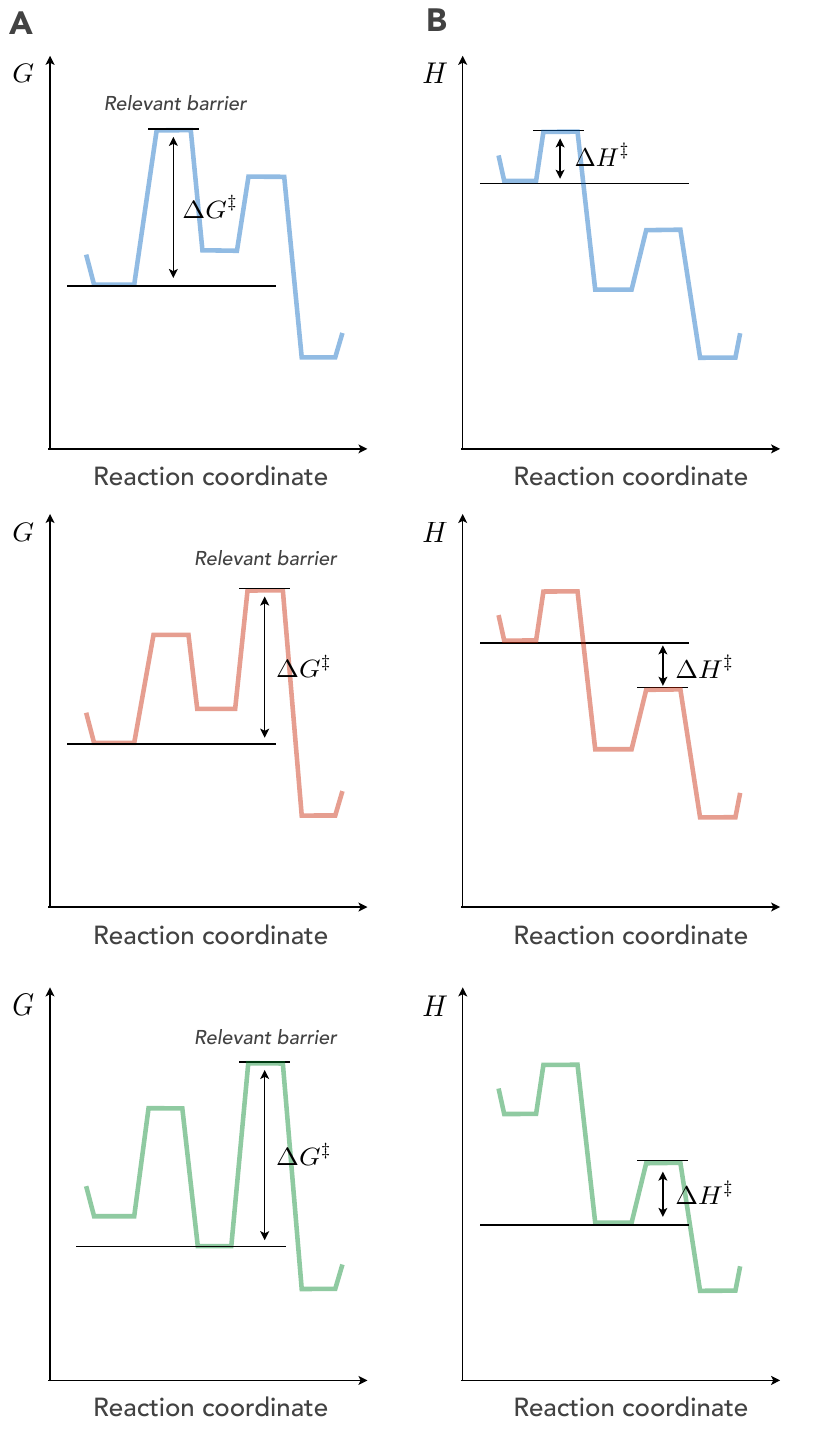}\end{center}
\caption{{\bf{Relating temperature-dependent measurements of reaction rates to the relevant barrier along the complex free energy landscape describing macromolecular processes.}} Eq.\eqref{rate2} is used to determine the relevant barrier for three examples (top, middle, bottom) of free energy (a) and enthalpy (b) landscapes.}
  \end{figure}

\subsection{Determining the relevant free energy barrier from kinetic experiments}

Using Eq.~\eqref{k} we find that the escape rate is in the form of the Arrhenius-Eyring equation:
\begin{equation}\label{rate}
k(x_0\to x_e)  \simeq A \ e^{-\beta  \Delta G^\ddag},
\end{equation}
where $A=\frac{\omega_1\omega_2}{2\pi \gamma}$ is a pre-factor, which depends on the curvatures of the potential landscape at $x_*$ and $x^*$, respectively. Note that, although the energy landscape includes multiple intermediate kinetic barriers, only one relevant free energy barrier $\Delta G^{\ddag}$ determines the escape rate $k(x_0\to x_e)$ and hence can be probed directly by kinetic experiments. This relevant free energy barrier is found using Eq.~\eqref{rate2} and corresponds to the largest possible energy difference between any local maximum and any local minimum preceding it. Figure 2 illustrates this principle for a series of three examples of energy landscapes. An analogous interpretation of the results can be given by performing an explicit analysis of the spectrum of the rate matrix that describes individual transitions in the energy landscapes of Fig.~2 (see Supplementary Material for details).

As Eq.~\eqref{rate} predicts that $\tau(x_0\to x_e)$ has an exponential dependency on the height of the relevant energy barrier, using the relationship $\Delta G^{\ddag} = \Delta H^{\ddag}-T \Delta S^{\ddag}$, where $\Delta H^{\ddag}$ is the enthalpy of activation and $\Delta S^{\ddag}$ is the entropy of activation, and absorbing the entropy contribution into the prefactor, we find $k(x_0\to x_e) \simeq e^{-\beta \Delta H^{\ddag}}$. Hence, a plot of $\log k(x_0\to x_e)$ against $\beta = 1/k_BT$ is expected to yield a straight line with the enthalpy of activation corresponding to the relevant barrier as the slope:
\begin{equation}\label{DH}
\frac{\Delta H^{\ddag}}{k_B} =-  \frac{\partial \log k(x_0\to x_e)}{\partial (1/T)}.
\end{equation}
This equation provides the key for interpreting reaction rate measurements at varying temperature in terms of the enthalpy of activation of the rate determining step. Note, however, that replacing $\Delta G^{\ddag}$ with $\Delta H^{\ddag}$ in Eq.~\eqref{rate2} is wrong. This is because, the relevant activation energy barrier is 
determined by the free energy landscape (Fig.~2a), while the measured 
temperature dependence of rate constants only reflects the enthalpic 
contribution to this barrier, which need not correspond to the highest 
enthalpy change (Fig.~2b).

To numerically test the theoretical predictions of Eqs.~\eqref{rate2}-\eqref{DH}, we performed Monte Carlo (MC) and Molecular Dynamics (MD) simulations ~\cite{Daans_book} of diffusion of a single particle on a series of examples of one-dimensional potential energy landscapes (see Supplementary Material).
 
\subsection{The frequency pre-factor}

Unlike the enthalpy of activation, $\Delta H^{\ddag}$, the free energy of activation, $\Delta G^{\ddag}$, is crucially coordinate dependent. This raises the question of the appropriate choice for the reaction coordinate, or, equivalently, of an appropriate frequency pre-factor $A$. While Kramers theory in principle provides an explicit formula for this pre-factor via Eq.~\eqref{rate}, this expression depends on parameters such as the curvature of the free energy landscape at the top of the relevant barrier, which are commonly inaccessible in experiments.
A possible strategy to overcome this limitation consists in partitioning all of the missing information about diffusion along the reaction coordinate into the free energy barrier in the rate equation by re-writing the escape rate as:
\begin{align}
k(x_0\to x_e)  & = A^{\rm{phys}} \ e^{-\beta \Delta G^{\ddag,\rm{phys}} },
\end{align}
where
$ \Delta G^{\ddag,\rm{phys}} = \Delta G^{\ddag}+ T\Delta S^{\rm{i}}$ and $\Delta S^{\rm{i}} = k_B\log(A^{\rm{phys}}/A)$.  Here, $A^{\rm{phys}}$ is a known frequency pre-factor which can be constructed conveniently from the experimentally accessible information about the reactive flux towards the relevant barrier. Hence, $\Delta S^{\rm{i}}$ can be interpreted as an additional entropy term representing the fact that we inevitably do not have complete information about diffusion along the reaction coordinate.  Note that various choices of partitionings are equally possible. Thus, the kinetic pre-factor, $A$, and the absolute value of the relevant free energy barrier, $\Delta G^{\ddag}$, are not independent quantities, and the latter is only meaningful if stated together with the corresponding pre-factor.

\section{Application to protein aggregation}  

\begin{figure*} 
\includegraphics[width=\textwidth]{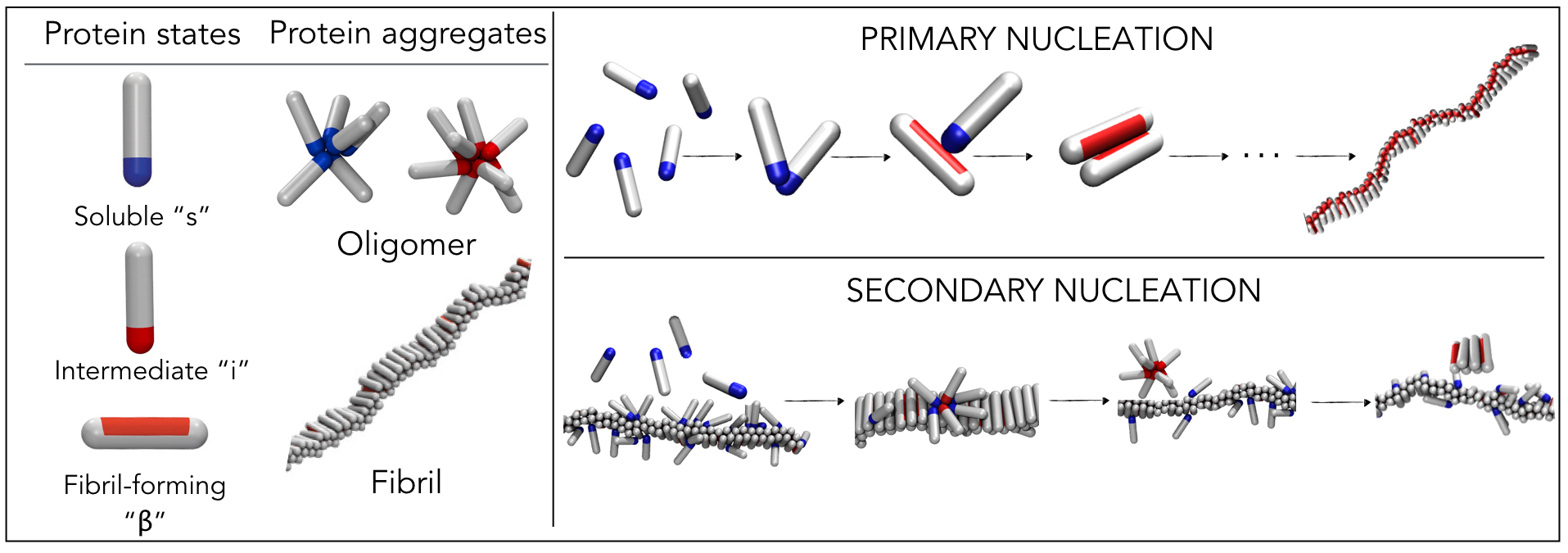}
\caption{\textbf{Coarse-grained model for amyloid aggregation}. (a) Monomers can exist in two states - the soluble state that forms oligomers, and a $\beta$-sheet-prone state that forms fibrils. When bound to a fibril, monomers can also convert into an intermediate state, which binds stronger to its own kind than to the fibril and hence self-associates into oligomers that detach from the fibril surface. (b) Primary nucleation, over the concentration and temperature regime discussed in this paper ($c=1.8$mM,$ 1.25 < k_BT <10 $), proceeds through protein dimerization and conversion into a $\beta$-sheet dimer which continues growing into a fully-elongated fibril. Secondary nucleation ($c=0.15$mM, $ 0.92 < k_BT <1.03 $) proceeds by monomer attachment and oligomerization on the fibril surface, conversion into an intermediate state, oligomer detachment, and finally conversion into the $\beta$-sheet rich nucleus in solution. }\label{fig_model}
 \end{figure*}

In the following, we demonstrate how Kramers rate theory discussed above can be used to study complex multi-molecular reactions governed by diffusive dynamics and, in this manner, extract energetic information about some of its constituents microscopic steps. In particular, we shall focus on protein aggregation kinetics into amyloid fibrils, a process which is attracting great interest due to its connection with over 50 medical conditions, including Alzheimer's and Parkinson's diseases \cite{Chiti_2006,Chiti_2017,Dobson_2017,Knowles_2014}. 

Amyloid fibril formation is a process in which soluble proteins spontaneously aggregate into fibrils of a cross-$\beta$ structure, enriched in $\beta$-sheet content~\cite{Fitzpatrick_2013}. This is a complex phenomenon that typically involves the concomitant action of multiple molecular mechanisms. Recent advances in the available experimental techniques for measuring aggregation kinetics coupled to mathematical analysis of the underlying kinetic equations have allowed the identification of these mechanisms at a microscopic level \cite{Michaels_2018,Cohen_2012,Meisl_2016}. In the case of the aggregation of A$\beta$42 (the 42-residue form of the Amyloid-$\beta$ peptide), a process that is intimately linked to Alzheimer's disease \cite{Selkoe_2016}, the fundamental steps that underlie amyloid fibril formation involve an initial primary nucleation step, where monomeric proteins spontaneously come together to form new fibrils, coupled to filament elongation. In addition, the aggregation process is accelerated by the fact that fibrils are able to generate copies of themselves through surface catalysis \cite{Cohen_2013,Meisl_2014}, a process known as secondary nucleation \cite{Ferrone_1985}.

Despite the fact that the molecular steps of A$\beta$42 amyloid formation have been identified, the molecular mechanisms that underlie them have remained challenging to understand \cite{Kashchiev_2010,Cabriolu_2010,Cabriolu_2011}. Here, we use Kramers theory to study the free energy landscape of two key steps in the formation of A$\beta$42 amyloid fibrils, namely primary and secondary nucleation. We use a coarse-grained model (Fig.~3) which can capture the diffusive motion of proteins on a multi-barrier energy landscape determined by the relevant effective intermolecular interactions, such as hydrophobic forces, hydrogen bonding and screened electrostatic interactions. Our model retains only crucial molecular ingredients needed to reproduce the aggregation behaviour at experimentally relevant scales. The main advantage of this coarse-grained computer simulation approach is that the measurements from simulations can be validated directly against bulk experimental measurements~\cite{Saric_2016}, as we also demonstrate here.\\ 
In the case of primary nucleation, the free energy landscape along the reaction coordinate in our model involves an initial step whereby monomeric proteins associate, followed by a conformational conversion of proteins from their native soluble states into a $\beta$-sheet prone state, and finally the association of the latter state in a $\beta$-sheet rich nucleus, which then rapidly grows into an amyloid fibril (Fig. 3). Secondary nucleation involves the adsorption of monomers onto the surface of an existing fibril and a subsequent surface-catalysed conformational conversion step. This step leads to aggregate detachment and its conversion into the $\beta$-sheet nucleus, which then elongates into a fibril (Fig. 3). In both scenarios, we measure the temperature dependence of the overall rates of primary and secondary nucleation, and show that, despite the complexity of the underlying processes, these temperature-dependent kinetic measurements probe only one relevant barrier along the free energy landscape. We then compare our simulation results to the equivalent experimental results on the temperature-dependence of primary and secondary nucleation during the formation of A$\beta$42 amyloid fibrils. Just like predicted by Eq.~\eqref{rate} and \eqref{DH}, we show that the kinetic measurements do not probe the highest enthalpic barrier on the energy landscape, but rather the enthalpic barrier which contributes tot the highest free energy barrier on the landscape.

\begin{figure*} 
\includegraphics[width=0.99\textwidth]{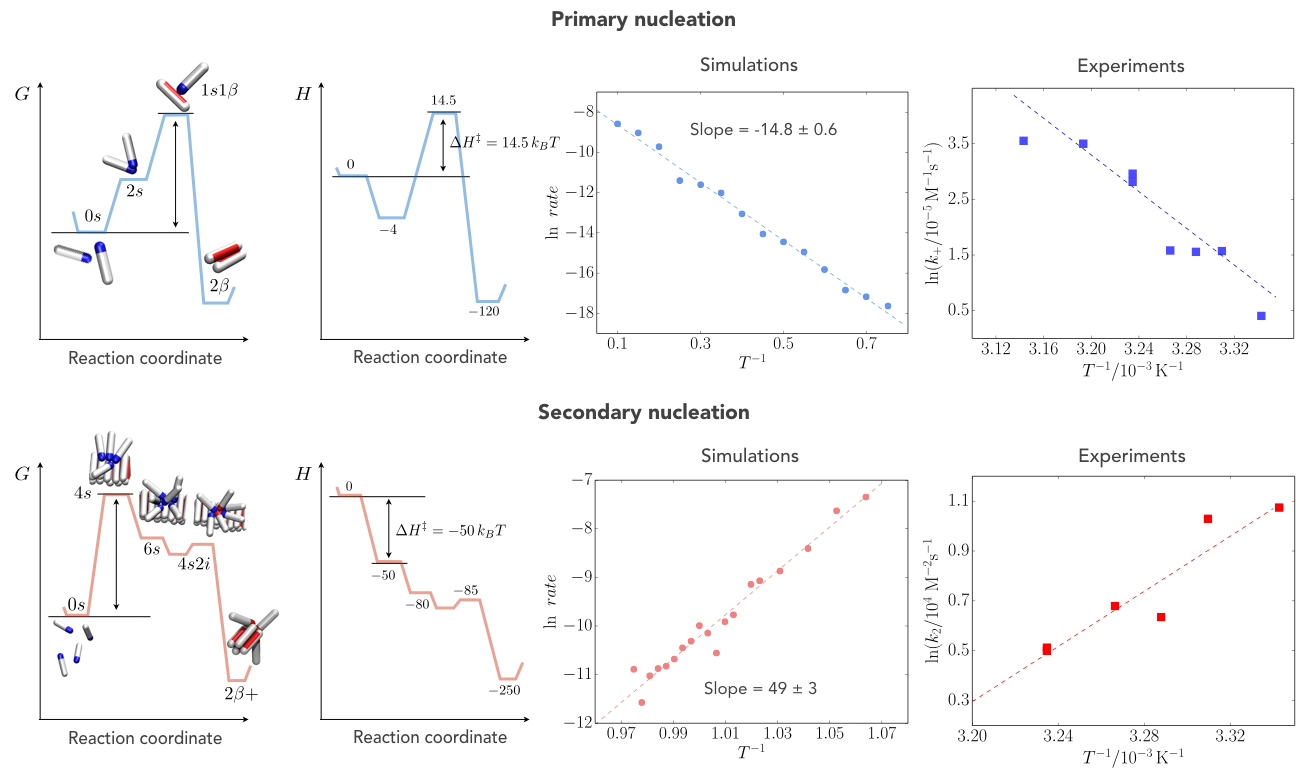}
\caption{{\bf{Determining the rate limiting step for primary and secondary nucleation in amyloid fibril formation.}} Free energy and enthalpy profiles underlying primary (top) and secondary (bottom) nucleation in our computer model and in experiments, and the variation of the respective nucleation rates with temperature. \textbf{Top panel:} The highest free energy barrier for primary nucleation in the simulations considered here corresponds to the conversion of a single monomer in the nucleus to the $\beta$-sheet configuration (left). The temperature dependence of the nucleation rate is readout of the energetic penalty for this conversion, imposed by $\Delta\mu_{s-\beta}$ in our model (middle). The experiments on A$\beta$42 peptide show the same trends in the temperature-variation of the primary nucleation rate (right). \textbf{Bottom panel:} The highest free energy barrier for secondary nucleation in this model corresponds to monomer adsorption onto the fibril (left). The variation of the nucleation rate with temperature then probes the enthalpic barrier for protein adsorption, which is in this case negative (middle). The experiments on secondary nucleation of A$\beta$42 exhibit the same trend~\cite{NatChem_2018}, with the nucleation rate being decreased at high temperatures (right). }
 \end{figure*}

\subsection{Computer model}
We used a coarse-grained Monte Carlo model for primary and secondary amyloid nucleation developed in \cite{Saric_2014} and \cite{Saric_2016}. Although minimal in its nature, this model captures many complex features of the aggregation processes. In particular, the model accounts for the fact that an amyloidogenic protein needs to exist in at least two different states: a state in solution (``$s$'') that can occasionally aggregate into small oligomers, and a higher free-energy state that can form amyloid fibrils (``$\beta$'')~\cite{vacha,Saric_2014} (Fig. 3). To capture secondary nucleation, a soluble monomer can adsorb onto the fibril surface, and can convert into an additional (intermediate ``$i$'') state that lies inbetween the ``$s$'' and ``$\beta$'' states; the existence of this intermediate state reflects the catalytic role of the fibril in assisting the conformational conversion from the soluble into the fibril-forming state. In this model a protein is described by a single rod-like particle, decorated with a patch that controls protein aggregation into either non-$\beta$-sheet oligomers or fibrils. A protein in an  ``$s$'' or ``$i$'' state interacts with its own kind via its attractive tip, which mimics non-specific inter-protein interactions. The fibril-forming state interacts with its own kind via an attractive side-patch, which models directional interactions, such as hydrogen bonding, and drives the formation of fibrillar aggregates. The interaction between two proteins in the fibril-forming states is by far the strongest interaction in the system, and once formed, the fibrils are effectively irreversible.  Monomer adsorption onto the fibril is energetically favorable; adsorbed monomers are then able to interact on the fibril surface to form oligomers. Since the protein in the intermediate state interacts with the fibril only weakly, oligomer detachment is favorable only for sufficiently large oligomers. This is because the loss of monomer-fibril interactions is overcome by the energetic gain associated with the interaction between proteins in the oligomer-forming state. Every conversion event from the soluble state into the fibril-forming state is penalized by a change in the excess chemical potential, $\Delta\mu_{s-\beta}$. This property is needed to reflect the fact that amyloidogenic proteins, such as A$\beta$, are typically not found in the $\beta$-sheet prone conformation in solution ~\cite{dobson3,vendruscolo}. 
As in our previous work \cite{Saric_2016},  the conversion from the soluble to the intermediate state on the fibril surface, as well as the conversion from the oligomer-forming state to the fibril forming state was penalized by $0.5\Delta\mu_{s-\beta}$. 
Further details on the parameters used in this work are given in the Supplementary Information. To calculate the nucleation rate, we measure the mean first passage time for the particles to diffuse along the energy landscape and create a $\beta$-sheet enriched nucleus. The rate of nucleation is then defined as the inverse of such an  average first passage time for nucleation~\cite{Saric_2016, JCP_2016}. \par

 \begin{figure*} 
\includegraphics[width=0.99\textwidth]{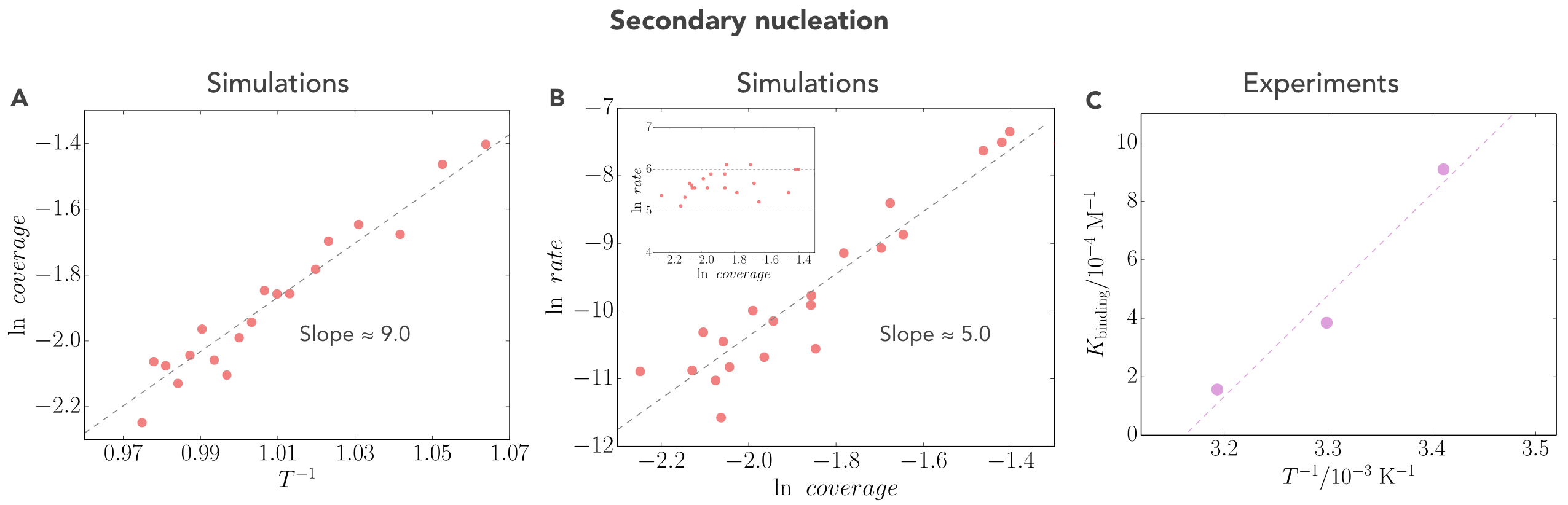}
\caption{{\bf{Temperature dependence of secondary nucleation.}} (a) Computer simulations show that the amount of monomers adsorbed on fibrils decreases with increasing temperature. (b) The rate of secondary nucleation in computer simulations depends linearly on the fibril surface coverage, while the size of the nucleating oligomer remains unchanged (Inset). The combination of the dependence in (a) and (b) results in the temperature-dependence of secondary nucleation presented in Fig. 4. (c)  The temperature dependence of the monomer-fibril binding constant measured for the A$\beta$ peptide shows the same trend as the temperature dependence of surface coverage observed in simulations ~\cite{NatChem_2018}.}
 \end{figure*}

\subsection{Temperature dependence of primary nucleation}
For primary nucleation to take place, proteins need to meet in solution and then convert into the $\beta$-sheet-prone conformations. The converted monomers interact strongly and form a $\beta$-sheet nucleus which is able to grow into a mature fibril (Fig. 3). This process can involve the formation of long-lived pre-nucleation clusters. Such clusters provide a suitable environment for the conformational conversion to take place and hence significantly enhance the rate of nucleation~\cite{Saric_2014, JCP_2016}. We simulated amyloid nucleation in solution for a range of different temperatures. Interestingly, we find a non-monotonic dependence of the primary nucleation rate on temperature (Fig. S5 a): at low temperatures the nucleation rate decreases with temperature, while at high temperatures the rate increases as the temperature is increased. We find that monomers oligomerize substantially at low temperatures (Fig. S5 b), which increases the rate of fibril nucleation. As the temperature is increased, the pre-nucleation oligomers become smaller and smaller, which in turn decreases the nucleation rate. However, as the temperature is further increased, the high-energy $\beta$-sheet prone state becomes more easily accessible by thermal fluctuations, and the conversion rate is enhanced, resulting in an increased overall nucleation rate. \\
Recently, the temperature-dependence of the rate of primary nucleation of  A$\beta$ peptide has been probed in experiments~\cite{NatChem_2018}, and it has been found that, in the experimentally relevant regime of temperatures for this peptide, the nucleation rate is significantly increased at higher temperatures. Hence we focus on this regime in our analysis. In this temperature regime the nucleation process starts by two proteins meeting in solution and forming a dimer (denoted as "$2s$" in Fig. 4). A "$2s$" dimer usually falls apart many times and reforms elsewhere in solution before one protein in the dimer successfully converts into the $\beta$-sheet prone state ("$1s1\beta$" in Fig. 4). Such a "$1s1\beta$" dimer also has a high probability of dissolving back to the solution, before a successful conversion into a $\beta$-sheet nucleus ("2$\beta$") occurs, which then quickly grows into a fibril. From analyzing the simulation trajectories, we find that such a nucleus made of two proteins in the $\beta$-sheet prone state always grows further. We also detect a significant amount of dimers which contain two proteins in the soluble states, while the least probable species in the system is a dimer that contains exactly one protein in the soluble state and one protein in the $\beta$-sheet prone state. We assign the highest free energy barrier in the system to precisely this rare species, as shown in Fig. 4. Since we know all the interactions in the system, we can explicitly calculate the enthalpic barrier which corresponds to this "rate-determining" free energy barrier in our simulations. Our measurements recover the value of 14.5 $k_BT$ for the relevant enthalpic barrier, which agrees remarkably well with the variation of the reaction rate with temperature that measures 14.8 $k_BT$, just like predicted by the reaction rate theory in Eqs.(\ref{rate}-\ref{DH}).

\subsection{Temperature dependence of secondary nucleation}
We repeated an analogous set of simulations and rate measurements for the temperature dependence of secondary nucleation. By contrast to primary nucleation, secondary nucleation occurs via protein adsorption and oligomerization on the fibril surface, followed by a conformational conversion into the intermediate state, oligomer detachment, and conversion into the $\beta$-sheet nucleus which further grows into a fibril in solution (Fig. 3). From our simulations, at the particular protein concentration we considered, we find that the rarest species in the system is a fibril-bound oligomer which contains four proteins in the "s" state ("$4s$" in Fig.4). If such an oligomer survives, it grows into a hexamer ("$6s$" in Fig. 4), which then partially converts into an intermediate state ("$4s2i$" in Fig.4), detaches ("$6i$" in Fig.4), and finally converts into a $\beta$-sheet nucleus ("$4i2\beta$" in Fig.4).  Our rate measurements show that secondary nucleation is strongly hampered at high temperatures, which is exactly the opposite trend compared to the one observed for primary nucleation. The reason for this observation is that, unlike in primary nucleation, the highest free energy barrier for secondary nucleation corresponds to the protein oligomerisation step on the fibril surface, rather than the oligomer conversion step. At higher temperatures, fewer monomers are adsorbed onto the fibril surface, which leads to a decreased protein oligomerisation, and slower overall nucleation.
 It is important to note that secondary nucleation in our simulations occurs only in a very narrow temperature regime. As previously reported in experiments~\cite{Meisl_2014, Meisl_2017} and simulations~\cite{Saric_2016}, secondary nucleation is extremely sensitive to environmental conditions. The exact temperature range where secondary nucleation occurs in our simulations is determined by the choice of all the interactions in the system, which is somewhat arbitrary in such a highly coarse-grained model. Hence, one should not try to compare the exact values of the rates or temperatures between primary and secondary nucleation, but should rather focus on trends and qualitative behavior observed in simulations. \\
To that end, Fig. 5a shows the temperature-dependence of the surface coverage, while Fig. 5b plots the dependence of the rate of secondary nucleation on the fibril surface coverage. The two dependencies (Fig. 5a and b) combined give rise to the temperature-dependent behaviour of the secondary nucleation reaction rate observed in Fig. 4. From the measurements of the interactions in our model (Fig. 4), we find that the enthalpic barrier corresponding to the formation of the "critical" oligomer is actually negative, and, again, this result aligns very well with the variation of the nucleation rate with temperature. The same trends have recently been reported for the kinetics of secondary nucleation of A$\beta$42~\cite{NatChem_2018}, where the secondary nucleation has also been reported to be retarded at high temperatures, and appeared to have a negative enthalpic barrier (Fig. 4). In this case, the observed behaviour is also caused by the lower protein-fibril adsorption as the temperature increases~\cite{NatChem_2018} (Fig. 5c). Secondary amyloid nucleation is a clear example of a process in which the kinetic measurements do not read out the highest enthalpic barrier on the energy landscape. The enthalpic barrier probed is highly negative, both in simulations and experiments, and contributes to the highest free energy barrier on the free energy landscape, which is clearly dominated by the unfavourable entropic contribution related to protein adsorption.

\section{Conclusions}
We have discussed a general framework, based on Kramers reaction rate theory, for studying the temperature-dependence of complex supramolecular processes with multiple barriers and wells. We find that the enthalpic barrier probed in temperature dependent kinetic measurements of reaction rates does not necessarily correspond to the highest enthalpic barrier along the reaction coordinate, but rather to the enthalpic barrier corresponding to the highest relative free energy barrier on the free energy landscape. We have then applied Kramers theory to interpret in coarse-grained computer simulations of the fundamental processes underlying the formation of amyloid fibrils - primary and secondary amyloid nucleation. For primary nucleation we find that two regimes can exist for the rate of nucleation - a regime in which the nucleation is faster at low temperatures, and a regime in which nucleation is faster at high temperatures. Guided by recent experimental results, we focus on the latter regime, and find that protein conformational conversion, which is aided at high temperatures, is driving primary nucleation in this temperature range. Unlike primary nucleation, we find that the relevant free energy barrier that determines the temperature dependence of secondary nucleation is the adsorption and oligomerisation on the fibril surface, which is hampered at high temperature. This difference results in fundamentally distinct thermodynamic signatures for primary and secondary nucleation, thus highlighting the power of probing free energy landscapes for understanding microscopic processes underlying complex multi-molecular processes.

\section*{Acknowledgements}

LXL ans SC contributed equally to this work. We acknowledge support from the Swiss National Science Foundation (TCTM), Peterhouse College Cambridge (TCTM), the UCL Institute for the Physics of Living Systems (LXL, SC, A\v{S}), ERASMUS Placement Programme (SC), the Royal Society (A\v{S}), the Academy of Medical Sciences (A\v{S}), Wellcome Trust (A\v{S}), the Biological Sciences Research Council (TPJK), the Frances and Augustus Newman Foundation (TPJK), the European Research Council (TPJK). We thank Claudia Flandoli for the help with illustrations.



\section*{Supplementary material}

# Supplementary Material to "Reaction rate theory for supramolecular kinetics: application to protein aggregation."

March 13, 2018

## 1 Kramers theory of diffusive reactions with multiple kinetic barriers

In this section, we present a detailed derivation of some key equations of Kramers theory of diffusive reactions with multiple kinetic barriers, which are discussed in Section II of the main text.

Let us consider a Brownian particle moving in one-dimensional potential landscape $G(x)$, such as the one sketched in Fig. 1 of the main text. It is assumed that the mean thermal energy of the Brownian particle is much smaller than the free energy barrier height. The Brownian particle starts diffusing from an initial location $x_0$ lying somewhere in some region of space $\Omega$ with boundary $\partial\Omega$. In the following we will consider the case when $\Omega = (-\infty, x_e]$. The key question that we want to address is: how long does it take on average for the Brownian particle to diffuse from $x_0$ to the boundary $\partial\Omega$ and hence leave the set $\Omega$? The starting point for answering this question is the Fokker-Planck equation, which describes the time evolution of the probability $p(x, t|x_0)$ that the Brownian particle will be at position $x$ at time $t$ given that

it started at position $x_0$ at $t = 0$. The Fokker-Planck equation reads (Eq. (1) of the main text):

$$\frac{\partial}{\partial t}p(x,t|x_0) = \frac{\partial}{\partial x}\left[\frac{1}{\gamma}\frac{\partial G(x)}{\partial x}p(x,t|x_0)\right] + D\frac{\partial^2}{\partial x^2}p(x,t|x_0) \tag{S1}$$

where $\gamma$ is the frictional coefficient (note: we assume to be in the large friction limit so that the inertia term can be neglected), $D$ is the diffusion coefficient and $\gamma D = k_B T$.

The key quantity of interest is the first passage time $\tau(x_0)$, defined as the first time at which a Brownian particle starting at $x_0$ hits the point $x_e$ defining the boundary of $\Omega$. Since $\tau(x_0)$ is a random variable, we consider the probability distribution $T(t|x_0)$ that $\tau(x_0)$ equals $t$. According to the theory of first passage times [1], $T(t|x_0)$ is computed as follows

$$T(t|x_0) = -\frac{dQ(t|x_0)}{dt}, \tag{S2}$$

where $Q(t|x_0)$ is the probability that at time $t$ a Brownian particle starting at $x_0$ has not yet reached the boundary $\partial\Omega$, i.e.

$$Q(t|x_0) = \int_\Omega p(x,t|x_0)dx. \tag{S3}$$

To prove the equation $T(t|x_0) = -\frac{dQ(t|x_0)}{dt}$, we note that $Q(t|x_0)$ can be also interpreted as the probability that the first passage time is larger than $t$, that is $Q(t|x_0) = \text{Prob}[\tau > t]$. Hence:

$$Q(t|x_0) = \int_t^\infty T(s|x_0)ds. \tag{S4}$$

Taking the derivative with respect to $t$ on both sides, yields

$$\frac{dQ(t|x_0)}{dt} = -T(t|x_0), \tag{S5}$$

which is what we wanted to prove. The average first passage time is therefore given by (integration by parts)

$$\begin{aligned}\langle \tau(x_0)\rangle &= \int_0^\infty t\, T(t|x_0)dt = -\int_0^\infty t\,\frac{dQ(t|x_0)}{dt}dt \\ &= \int_0^\infty Q(t|x_0)dt = \int_0^\infty dt \int_\Omega dx\, p(x,t|x_0).\end{aligned} \tag{S6}$$

This is Eq. (3) of the main text. We now combine Eq. (S6) with Eq. (S1) to obtain a differential equation for $\langle \tau(x_0) \rangle$, which is given by:

$$-\frac{1}{\gamma} \frac{dG(x_0)}{dx_0} \frac{d\langle \tau(x_0) \rangle}{dx_0} + D \frac{d^2 \langle \tau(x_0) \rangle}{dx_0^2} = -1 \tag{S7}$$

subject to the following boundary condition

$$\langle \tau(x_0) \rangle = 0 \text{ for } x_0 \in \partial\Omega, \text{ i.e. } \langle \tau(x_e) \rangle = 0. \tag{S8}$$

To prove Eq. (S7), we first notice that if $p(x,t|x_0,t_0)$ satisfies the forward Fokker-Planck equation $\partial_t p(x,t|x_0,t_0) = L_{FP}(x) p(x,t|x_0,t_0)$ with respect to the variables $(x,t)$, with the Fokker-Planck operator

$$L_{FP}(x) = \frac{\partial}{\partial x} \left[ \frac{1}{\gamma} \frac{\partial G(x)}{\partial x} \right] + D \frac{\partial^2}{\partial x^2}, \tag{S9}$$

then $p(x,t|x_0,t_0)$ satisfies the backward Fokker-Planck equation with respect to the variables $(x_0, t_0)$, i.e.

$$\frac{\partial}{\partial t_0} p(x,t|x_0,t_0) = -L^\dagger_{FP}(x_0) p(x,t|x_0,t_0) \tag{S10}$$

where the adjoint Fokker-Planck operator is

$$L^\dagger_{FP}(x_0) = -\frac{\partial}{\partial x_0} \left[ \frac{1}{\gamma} \frac{\partial G(x_0)}{\partial x_0} \right] + D \frac{\partial^2}{\partial x_0^2}. \tag{S11}$$

Thus, applying the adjoint Fokker-Planck operator to $\langle \tau(x_0) \rangle$, we find

$$L^\dagger_{FP}(x_0) \langle \tau(x_0) \rangle = \int_0^\infty dt \int_\Omega dx \, L^\dagger_{FP}(x_0) p(x,t|x_0,t_0) \tag{S12}$$

$$= -\int_0^\infty dt \int_\Omega dx \, \frac{\partial}{\partial t_0} p(x,t|x_0,t_0),$$

where in the last step we used the backward Fokker-Planck equation (S10). Using the stationarity condition $p(x,t|x_0,t_0) = p(x,t-t_0|x_0,0)$, we can rewrite the above equation as

$$L^\dagger_{FP}(x_0) \langle \tau(x_0) \rangle = -\int_0^\infty dt \int_\Omega dx \, \frac{\partial}{\partial t_0} p(x,t|x_0,t_0) \tag{S13}$$

$$= \int_0^\infty dt \int_\Omega dx \, \frac{\partial}{\partial t} p(x,t|x_0,0) = -1,$$

which is Eq. (S7).

To solve Eq. (S7), we multiply on both sides by $e^{-\beta G(x_0)}$, yielding

$$\left(\frac{d}{dx_0}e^{-\beta G(x_0)}\right)\frac{d\langle\tau(x_0)\rangle}{dx_0} + e^{-\beta G(x_0)}\frac{d^2\langle\tau(x_0)\rangle}{dx_0^2} = -\frac{1}{D}e^{-\beta G(x_0)} \tag{S14}$$

This equation can be rewritten as

$$\frac{d}{dx_0}\left(e^{-\beta G(x_0)}\frac{d\langle\tau(x_0)\rangle}{dx_0}\right) = -\frac{1}{D}e^{-\beta G(x_0)} \tag{S15}$$

Integrating once, we find

$$e^{-\beta G(x_0)}\frac{d\langle\tau(x_0)\rangle}{dx_0} = -\frac{1}{D}\int_{-\infty}^{x_0} e^{-\beta G(z)}dz. \tag{S16}$$

Then we integrate another time using the boundary condition $\langle\tau(x_e)\rangle = 0$ and

$$\langle\tau(x_0)\rangle = -\frac{1}{D}\int_{x_e}^{x_0} e^{\beta G(y)}\int_{-\infty}^{y} e^{-\beta G(z)}dz. \tag{S17}$$

The solution to Eq. (S7) with the boundary condition (S8) is therefore:

$$\langle\tau(x_0)\rangle = \frac{1}{D}\int_{x_e}^{x_0} dy \int_{-\infty}^{y} dz\, e^{\beta[G(y)-G(z)]}. \tag{S18}$$

This is Eq. (6) in the main text. The integrals in Eq. (S18) can be evaluated using the saddle point approximation or Laplace's method [3]. To this end, we need to maximize the integrand over the integration range of Eq. (S18), i.e. we need to find $\max_{z\leq y}[G(y)-G(z)]$. Let us denote the points in the range of integration of Eq. (S18) where the integrand in Eq. (S18) is maximal as $y = x^*$ and $z = x_*$, and so using the following expansions (note that $G''(x^*) < 0$ since $x^*$ is a maximum point):

$$G(y) = G(x^*) - \frac{|G''(x^*)|}{2}(y-x^*)^2 + \text{ higher order terms} \tag{S19}$$

and

$$G(z) = G(x_*) + \frac{G''(x_*)}{2}(z-x_*)^2 + \text{ higher order terms}, \tag{S20}$$

S4

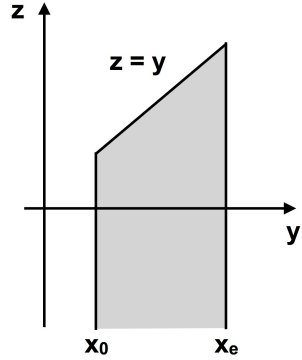

Figure S1: Schematic representation of the region of integration in Eq. (S18). To apply Laplace's method [3] on the integral in Eq. (S18), the function $G(y) - G(z)$ must be maximized on this region, leading to Eq. (S23).

we find:

$$\langle \tau(x_0) \rangle = \frac{2\pi}{\beta D \sqrt{|G''(x^*)|\ G''(x_*)}} e^{\beta[G(x^*) - G(x_*)]}, \tag{S21}$$

where we have used the result for Gaussian integrals $\int_{\mathbb{R}} e^{-\alpha x^2} dx^2 = \sqrt{\pi/\alpha}$ for $\alpha > 0$. The transition rate from $x_0$ to $x_e$ is the inverse of the average first hitting time, $k(x_0 \to x_e) = 1/\langle \tau(x_0) \rangle$, and so, using $\beta D = 1/\gamma$, we find:

$$k(x_0 \to x_e) = \frac{\sqrt{|G''(x^*)|\ G''(x_*)}}{2\pi\gamma} e^{-\beta \Delta G^{\ddagger}} = \frac{\omega_1 \omega_2}{2\pi\gamma} e^{-\beta \Delta G^{\ddagger}}, \tag{S22}$$

where

$$\Delta G^{\ddagger} = G(x^*) - G(x_*) = \max_{x_0 \leq z \leq y \leq x_e} [G(y) - G(z)], \tag{S23}$$

and $\omega_1$ and $\omega_2$ are the curvatures (i.e. second derivatives) of the potential at $x_*$ and $x^*$, respectively.

## 2 Details on MC and MD simulations

### 2.1 Diffusion of a single particle on a free energy landscape

To test the validity of key results Eqs. (S22) and (S23) (which are Eqs. (8) and (9) in the main text), numerical evidence is collected from Monte Carlo simulations of a particle moving in a

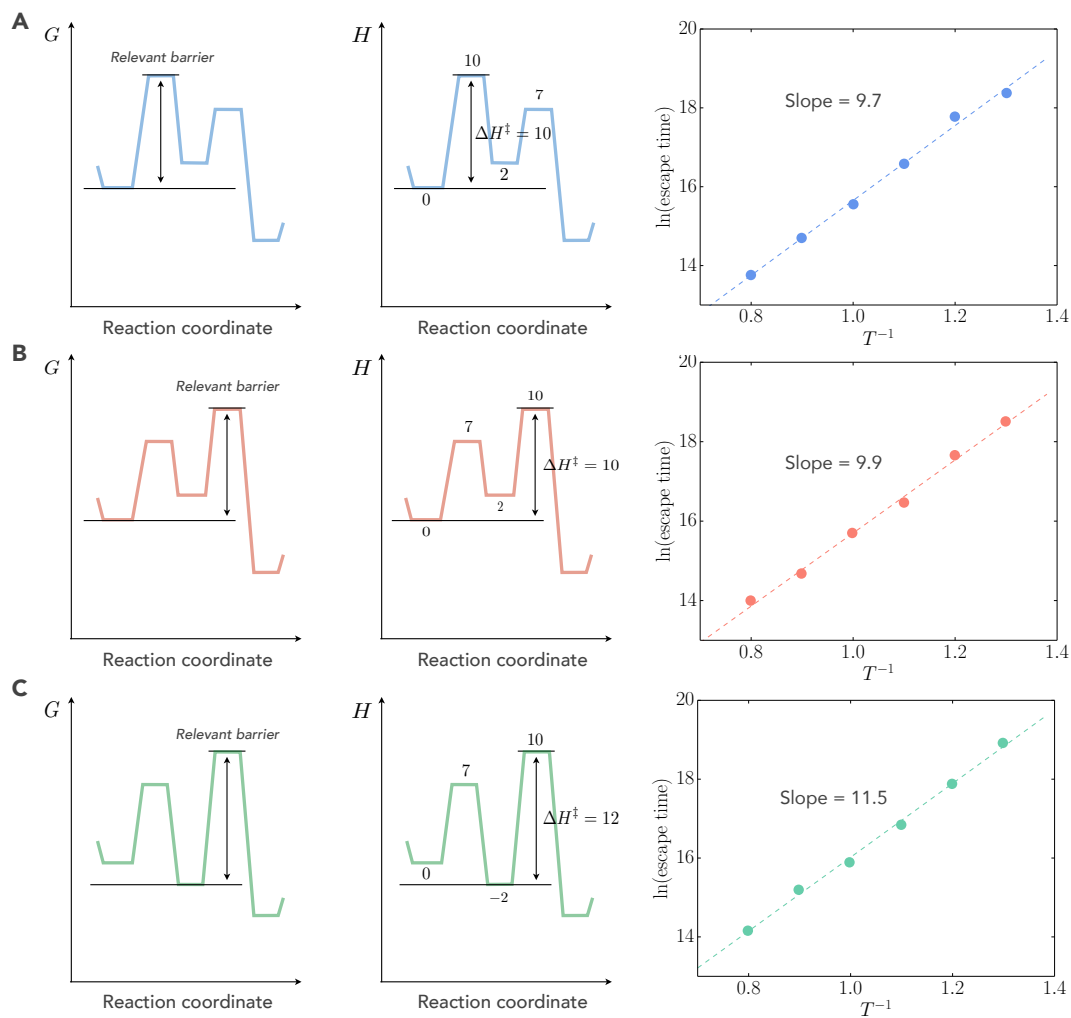

Figure S2: Eq. (S23) is used to determine the relevant barrier for three examples (a,b and c) of free energy (left) and enthalpy (middle) landscapes and comparison to MC computer simulations of the escape times, measured as the average number of MC steps needed o reach the global potential minimum (right).

one-dimensional box. The probability of acceptance of moves obey the Metropolis-Hastings algorithm, reflecting the Brownian particle's sensitivity to the surrounding energy landscape $\Delta G(x)$.

The particle is initially placed at the centre of the leftmost well and displaced randomly every step. To simulate diffusive processes under various temperatures, the width of the uniform step size distribution is set to be proportional to temperature, as described by Einstein's relation $D = k_B T/\gamma$. The average number of steps taken to reach the rightmost well for the first time is equivalent to $\langle \tau(x_0) \rangle$. As Equation (S22) states that $\langle \tau(x_0) \rangle$ depends exponentially on the relevant barrier height, an Arrhenius plot of $\log(\langle \tau(x_0) \rangle)$ against $\beta = 1/k_B T$ is expected to yield a straight line with the barrier height as the slope.

The first set of landscapes tested consist of three square wells separated by two square barriers defined on the interval $x \in [0, 1]$, with box boundaries $G(x) \to \infty$ for $x < 0$ and $x > 1$. Although unrealistic, simulations of square landscapes are particularly useful because on these landscapes, exact solutions to Eq.(S1) is obtainable through analytical methods, and can be easily compared to reflect accuracy of the results from simulation and test soundness of the simulation setup. It should be stressed that the theoretical results only become exact in the high barrier limit $G \gg k_B T$, but practically this condition is difficult to satisfy due to an exponential increase in CPU time with barrier height. This introduces a "systematic error" to the $\Delta G^\ddagger$ extracted from the Arrhenius plot, though it is in fact a property intrinsic to Eq.(S1)'s exact solution that reflects thermal effects of finite temperature. It is also expected that the limited number of crossings sampled results in a random spread in the plot.

Another subtlety arises from the fact that the landscape explicitly imposed in the algorithm is an enthalpy landscape, whereas entropy changes are implicitly imposed by the widths of potential wells accessible to the particle. To connect these simulations with theory, we can simply rewrite Eq. (S22) using $G = H - TS$ and absorbing the entropy contribution into the

S7

prefactor, we obtain:

$$k(x_0 \to x_e) \sim e^{-\beta \Delta H^\ddagger}, \tag{S24}$$

where the activation enthalpy $\Delta H^\ddagger$ is the quantity measured from simulations. However, the analogous statement:

$$\Delta H^\ddagger = H(x^*) - H(x_*) = \max_{x_0 \leq z \leq y \leq x_e} [H(y) - H(z)] \tag{S25}$$

is not true because the criterion for determining the activation energy still follows Eq. (S23). Although reactions on the landscapes tested in this work do not involve any significant entropic change, it is not uncommon to find condensed phase reactions with enthalpy and free energy landscapes so drastically different that the height ordering of minima and maxima becomes different. Any potential confusion may be resolved by realizing that the relevant activation energy barrier is determined by the free energy landscape, while the measured temperature dependence of rate constants only reflect the enthalpic contribution to this barrier. In other words, the relevant barrier is found by using Eq. (S23) on the free energy landscape; the measured $\Delta H^\ddagger$ is then the enthalpy change associated with the relevant free energy barrier; $\Delta H^\ddagger$ need not correspond to the highest enthalpy change. The same applies to experimental measurements of real chemical reactions (see e.g. Figure 4 of the main text).

Since physical energy landscapes of reactions are usually smooth rather than square, we moved on to test the reaction rates on a set of smooth landscapes, in which case the external force on the diffusing particles are continuous and we may find the numerical solution of Eq. (S1) from Molecular Dynamics simulations. Additionally, the friction coefficient $\gamma$ is set explicitly in MD, which allows the transition between ballistic and diffusive regimes to be controlled. This is in contrast to MC which could only probe diffusive reaction due to the random walk nature of the MC moves.

The results of MD are shown in Fig. S3. We also find that the results in Eq. (S22) and

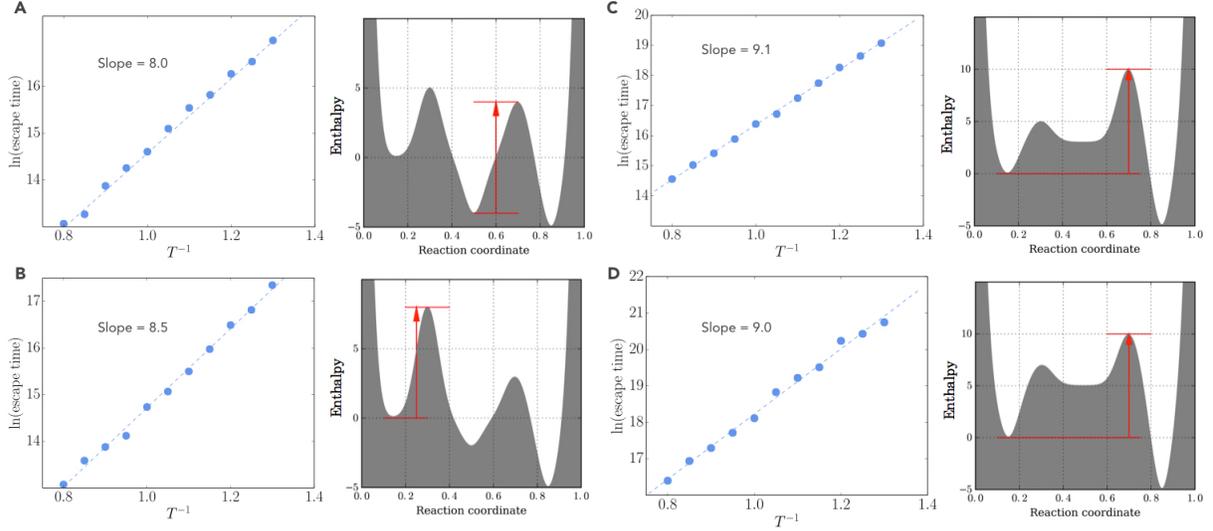

Figure S3: **Relevant barriers found from MD simulations on smooth landscapes:** red arrows mark the enthalpy barrier that enters the exponent of Arrehenius equation. Reaction starts from $x = 0.15$ and is recorded as finished when the diffusing particle crosses $x = 0.80$.

Eq. (S23) were indeed obeyed both in MC and MD when testing on the same smooth landscapes.

## 2.2 Coarse-grained model for amyloid nucleation

We used coarse-grained Monte Carlo model for primary and secondary amyloid nucleation developed in [6] and [7]. The model, potentials, and the notation are kept as in the original papers. The parameters of the model are summarised in Fig. S4. The exact parameters used in this work are as follows. For both primary and secondary nucleation: $\epsilon_{\beta\beta} = 60 k_B T$, $\epsilon_{s\beta} - \epsilon_{ss} = 1 k_B T$, and $\mu_{s->\beta} = 20 k_B T$. For primary nucleation: $\epsilon_{ss} = 4.5 k_B T$, and concentration of monomers $c = 1.8\,mM$. For secondary nucleation: $\epsilon_{ss} = 4 k_B T$, $\epsilon_{sf} = 6 k_B T$, $\epsilon_{i\beta} - \epsilon_{ii} = 1 k_B T$, $\epsilon_{if} = 1 k_B T$, $\epsilon_{si} = 8 k_B T$, $\epsilon_{ii} = 16 k_B T$, $\mu_{s->i} = 10 k_B T$, $\mu_{i->\beta} = 10 k_B T$, and the concentration of monomers $c \approx 0.15\,mM$.

## 2.3 Results of Coarse-grained simulations of amyloid aggregation

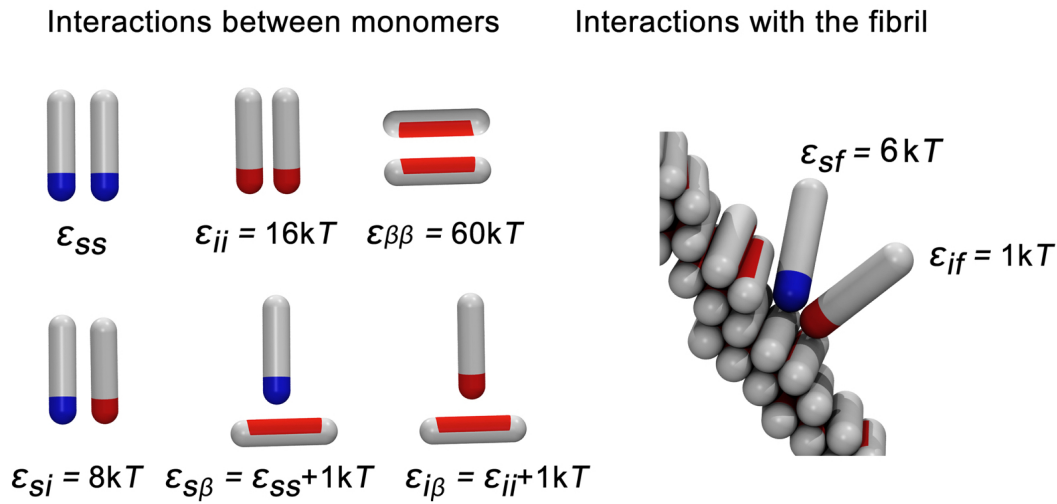

Figure S4: Coarse-grained model: possible interactions in the system.

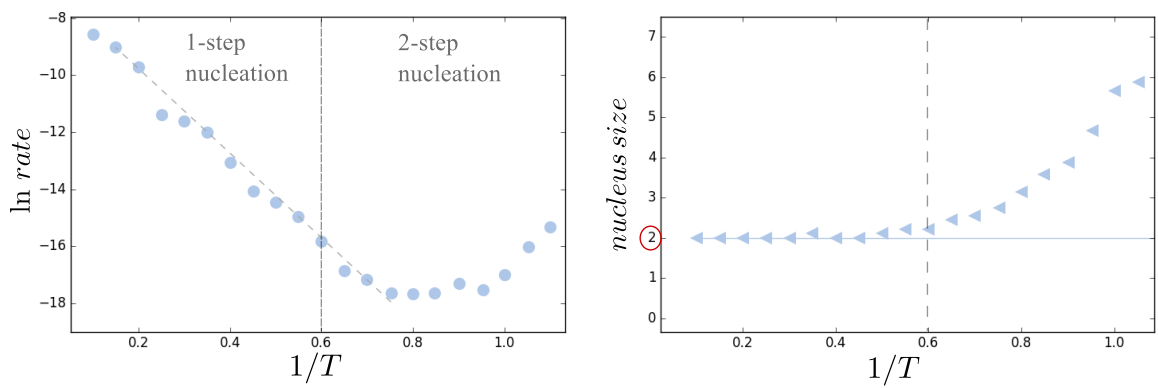

Figure S5: **Coarse-grained simulations of temperature dependence of primary nucleation.** (a) Two regimes are observed: a low-temperature regime, where the nucleation proceeds via stable pre-nucleation oligomers ("two-step nucleation"), and a high-temperature regime, where the nucleation proceeds by protein conversion either in solution or within small unstable dimers ("one-step nucleation"). (b) The size of the nucleating oligomer increases as the nucleating mechanism changes from high to low temperatures.

# 3 Analysis of three state kinetics using the master equation

## 3.1 Spectral analysis of the rate matrix

This section contains an explicit spectral analysis of the rate matrix for the kinetics in a three state system using a chemical master equation approach. The first part is mostly textbook

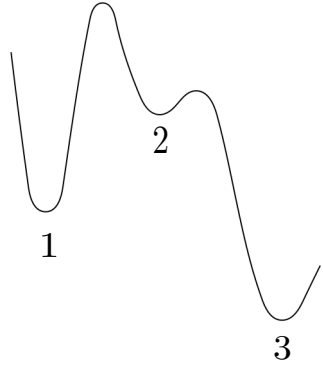

Figure S6: Schematic representation of three-state kinetics.

material, and can be found e.g. in van Kampen [4] or in Jackson [5]. The second part is specific for this work, and allows the establishment of the conditions for effective two state kinetics, the rate determining barrier, and the approximate effective rate based on this condition and barrier. The aim of this analysis is to provide an alternative view on the results using the Kramers' theory. Further, it allows a quick route to an analysis of all possible landscapes, not only the ones discussed in the Fig. 2 of the Main Text.

The (chemical) master equation for three subsequent states 1,2,3 depicted in Fig. S6 reads

$$\begin{pmatrix} \dot{p}_1 \\ \dot{p}_2 \\ \dot{p}_3 \end{pmatrix} = \begin{pmatrix} k_{21} p_2 - k_{12} p_1 \\ k_{12} p_1 + p_2(-k_{21} - k_{23}) + k_{32} p_3 \\ k_{23} p_2 - k_{32} p_3 \end{pmatrix}, \quad (S26)$$

where $p_1, p_2, p_3$ are respectively, the time dependent populations, and $k_{ij}$ denotes the kinetic rate constant between state $i$ and $j$. Note that $p_1 + p_2 + p_3 = 1$. This master equation can be conveniently re-written in matrix form as

$$\dot{\mathbf{p}} = \mathbf{K}\mathbf{p}, \quad (S27)$$

with

$$\mathbf{K} = \begin{pmatrix} -k_{12} & k_{21} & 0 \\ k_{12} & -k_{21} - k_{23} & k_{32} \\ 0 & k_{23} & -k_{32} \end{pmatrix} \quad (S28)$$

and $\mathbf{p} = \{p_1, p_2, p_3\}$. The solution of this equation is in general

$$\mathbf{p}(t) = \mathbf{p}(0) \exp(-\mathbf{K}t), \tag{S29}$$

where $\mathbf{p}(0)$ is the vector of initial conditions. Alternatively the solution can be expanded in eigenfunctions $\Phi_\lambda$ of $K$, which are given by [4]

$$\mathbf{K}\mathbf{\Phi_\lambda} = \lambda \mathbf{\Phi_\lambda}, \tag{S30}$$

so that the solution can be written as

$$\mathbf{p}(t) = \sum_\lambda c_\lambda \mathbf{\Phi_\lambda} e^{\lambda t}. \tag{S31}$$

The largest eigenvalue is $\lambda_0 = 0$ and its corresponding eigenvector is $\Phi_0 = \mathbf{p}_{eq}$. For the three state system the other two eigenvalues can be expressed analytically as

$$\begin{aligned}
\lambda_1 &= -\frac{1}{2}\left(k_{sum} - \sqrt{(k_{sum})^2 - 4(k_{12}k_{23} + k_{12}k_{32} + k_{21}k_{32})}\right) \\
\lambda_2 &= -\frac{1}{2}\left(k_{sum} + \sqrt{(k_{sum})^2 - 4(k_{12}k_{23} + k_{12}k_{32} + k_{21}k_{32})}\right),
\end{aligned} \tag{S32}$$

where $k_{sum} = k_{12} + k_{21} + k_{23} + k_{32}$.

The eigenvalues appear in the exponents of the solution, and can be identified with exponential relaxation processes. The relaxation time of those processes are given by $\tau = -1/\lambda$. The largest eigenvalue corresponds to infinite relaxation time, i.e. the equilibrium distribution. The other two relaxation times $\tau_1$ and $\tau_2$ depend on the specifics of the problem. When the times are similar to each other, the system exhibits typically three state kinetics, in which the processes interfere. In contrast, when the times are very disparate, the time scale separation make one relaxation process the dominate one, and the entire process can be described essentially by two-state kinetics with a single exponential relaxation time. For that to be the case we require that the fraction $\tau_2/\tau_1$ becomes smaller than a threshold $\alpha$, e.g. $\alpha \approx 0.1$. This threshold value

is a bit arbitrary, and should be in principle small, but in practice an order of magnitude in time scale can be considered sufficiently different to obtain effective two state kinetics.

We can thus write

$$\frac{\tau_2}{\tau_1} = \frac{\lambda_1}{\lambda_2} = \frac{1-\sqrt{1-4D}}{1+\sqrt{1-4D}} < \alpha, \tag{S33}$$

with

$$D = \frac{k_{12}k_{23} + k_{12}k_{32} + k_{21}k_{32}}{(k_{12} + k_{21} + k_{23} + k_{32})^2}. \tag{S34}$$

Up to now we made no assumption. To make progress we assume that $\alpha$ is sufficiently small to guarantee that $D$ is also small. Then we can expand the square root in Eq. S33 in terms of D, and can replace the condition $\frac{\tau_2}{\tau_1} < \alpha$ by

$$\frac{k_{12}k_{23} + k_{12}k_{32} + k_{21}k_{32}}{(k_{12} + k_{21} + k_{23} + k_{32})^2} < \alpha. \tag{S35}$$

Note that we would arrive at the same condition by first solving Eq. S33 for $D$ and then expand the r.h.s. in $\alpha$, and take the leading order.

This condition again depends on the specifics of the three state problem, e..g which rates are dominant in the system. The number of distinct possibilities is sufficiently large to refrain from doing an exhaustive analysis of this condition. However, to connect with the problem in the main text we can make the assumption, also made in the main text, that once the final state is reached, the reverse process is unlikely, i.e. that the process is almost irreversible. This means the $k_{32}$ is much smaller than any of the other rates. Hence, the above condition simplifies to

$$\frac{k_{12}k_{23}}{(k_{12} + k_{21} + k_{23})^2} < \alpha. \tag{S36}$$

Now we still have a number of possible cases. Suppose first that one single rate is dominant, i.e. much higher than the other two. Then we can distinguish three cases:

1) $k_{12}$ is dominant, i.e.. $k_{12} \gg k_{23}, k_{21}$ This would reduce the condition to

$$\frac{k_{12}k_{23}}{(k_{12})^2} = \frac{k_{23}}{k_{12}} < \alpha. \tag{S37}$$

S13

That is, the system would be effectively described by two-state kinetics if the rate over the second barrier is much slower then the rate over the first.

2) $k_{23}$ is dominant, i.e.. $k_{23} \gg k_{12}, k_{21}$ The condition is then

$$\frac{k_{12}k_{23}}{(k_{23})^2} = \frac{k_{12}}{k_{23}} < \alpha, \tag{S38}$$

the reverse of case 1). This would basically mean that the kinetics is completely determined by the rate over the first barrier. Thus, the system would be 2 state if the rate over the first barrier is much slower than the one over the second.

3). $k_{21}$ is dominant, e.. $k_{21} \gg k_{23}, k_{12}$, leading to

$$\frac{k_{12}k_{23}}{(k_{21})^2} < \alpha. \tag{S39}$$

Here there is not much direct insight to be gained.

The other extreme, namely that a single rate is much smaller than the other two, yields 3 additional cases: 4) First, we consider the case that $k_{12} \ll k_{23}, k_{21}$. The condition then becomes

$$\frac{k_{12}k_{23}}{(k_{21} + k_{23})^2} < \alpha. \tag{S40}$$

When the rates out of intermediate state 2 are about equal, i.e. $k_{23} \approx k_{21} \gg k_{12}$, this reduces to

$$\frac{k_{12}k_{23}}{(k_{21} + k_{23})^2} \approx \frac{k_{12}}{4k_{23}} < \alpha, \tag{S41}$$

very similar to case 2, but with an extra factor 4.

5) Of course we can also consider $k_{21} \ll k_{23}, k_{12}$, leading to

$$\frac{k_{12}k_{23}}{(k_{12} + k_{23})^2} < \alpha. \tag{S42}$$

When rates $k_{12} \approx k_{23}$, we obtain

$$\frac{k_{12}k_{23}}{(k_{12} + k_{23})^2} \approx \frac{1}{4} < \alpha, \tag{S43}$$

which is in fact, never fulfilled. Hence, for this case we can not expect effective two state kinetics under any circumstances.

6) Finally we can consider $k_{23} \ll k_{21}, k_{12}$, yielding,

$$\frac{k_{12}k_{23}}{(k_{12} + k_{21})^2} < \alpha \tag{S44}$$

which again, when assuming that $k_{12} \approx k_{21}$, simplifies to

$$\frac{k_{12}k_{23}}{(k_{12} + k_{21})^2} \approx \frac{k_{23}}{4k_{12}} < \alpha, \tag{S45}$$

very similar to case 1, but with an extra factor 4.

The above analysis is in terms of the individual rate constants of the 3 state process. However, as in the main text, one would also like expressions in terms of the free energies of the stable states and the barriers between them. For simplicity we express the rates in terms of free energies energy using the Arrhenius-Eyring expression

$$k_{ij} = \nu \exp(-\beta \Delta G_{ij}), \tag{S46}$$

where $\Delta G_{ij}$ denotes the barrier height for the transition from $i$ to $j$, and $\beta = 1/k_B T$ is the reciprocal temperature.

Denoting the (absolute) free energies of the stables state as $G_1, G_2, G_3$ and the absolute free energies of the barriers as $G_{12}^*$ and $G_{23}^*$ we have for the rate constants'

$$k_{12} = \nu \exp(-\beta(G_{12}^* - G_1)) \tag{S47}$$

$$k_{21} = \nu \exp(-\beta(G_{12}^* - G_2)) \tag{S48}$$

$$k_{23} = \nu \exp(-\beta(G_{23}^* - G_2)) \tag{S49}$$

$$k_{32} = \nu \exp(-\beta(G_{23}^* - G_3)), \tag{S50}$$

where we assume that all prefactors are identical. The condition for case 1) is now

$$\frac{k_{23}}{k_{12}} = \exp(-\beta(G_{23}^* - G_2))/\exp(-\beta(G_{12}^* - G_1)) < \alpha \tag{S51}$$

S15

which, taking the logarithm, turns into

$$-(G^*_{23} - G_2) + G^*_{12} - G_1 < k_B T \ln \alpha \tag{S52}$$

or

$$(G^*_{23} - G_2) - (G^*_{12} - G_1) > -k_B T \ln \alpha. \tag{S53}$$

Thus, the condition for 2-state kinetics is fulfilled when the second barrier is higher than the first by at least a few $k_B T$. For case 2) the situation is reversed:

$$\frac{k_{13}}{k_{23}} = \exp(-\beta(G^*_{12} - G_1))/\exp(-\beta(G^*_{23} - G_2)) < \alpha, \tag{S54}$$

which, taking the logarithm, turns into

$$(G^*_{12} - G_1) - (G^*_{23} - G_2) > -k_B T \ln \alpha, \tag{S55}$$

or the condition for 2-state kinetics is fulfilled when the first barrier is higher than the second by at least a few $k_B T$.

More complicated is the third case

$$\frac{k_{12} k_{23}}{(k_{21})^2} = \exp(-\beta(G^*_{12} - G_1)) \exp(-\beta(G^*_{23} - G_2))/\exp(-2\beta(G^*_{12} - G_2)) < \alpha, \tag{S56}$$

which, taking the logarithm, turns into

$$-(G^*_{12} - G_1)) - (G^*_{23} - G_2) + 2(G^*_{12} - G_2) < k_B T \ln \alpha. \tag{S57}$$

This condition can be simplified to

$$-G^*_{12} - G_1 + G^*_{23} + G_2 = (G^*_{23} - G_1) - (G^*_{12} - G_2) > -k_B T \ln \alpha, \tag{S58}$$

which states that 2 state kinetics is recovered when the second barrier height, measured with respect to state 1, is larger than the first barrier with respect to state 2, by at least a few $k_B T$.

| case | rate condition | 2-state condition | 2-state barrier condition | effective rate $k_f$ | landscape |
|---|---|---|---|---|---|
| 1 | $k_{12} \gg k_{23}, k_{21}$ | $\frac{k_2}{k_{12}} < \alpha$ | $(G^*_{23} - G_2) - (G^*_{12} - G_1) \gtrapprox 2. \, k_B T$ | $k_{23}$ | |
| 2 | $k_{23} \gg k_{12}, k_{21}$ | $\frac{k_{12}}{k_2} < \alpha$ | $(G^*_{12} - G_1) - (G^*_{23} - G_2) \gtrapprox 2. \, k_B T$ | $k_{12}$ | |
| 3 | $k_{21} \gg k_{23}, k_{12}$ | $\frac{k_{12} k_2}{(k_{21})^2} < \alpha$ | $(G^*_{23} - G_1) - (G^*_{12} - G_2) \gtrapprox 2. \, k_B T$ | $\frac{k_{12} k_2}{k_{21}}$ | |
| 4 | $k_{12} \ll k_{23} \approx k_{21}$ | $\frac{k_{12}}{k_2} < \alpha$ | $(G^*_{12} - G_1) - (G^*_{23} - G_2) \gtrapprox .2 k_B T$ | $\frac{k_{12} k_2}{k_{21} \, k_2} \approx \frac{1}{2} k_{12}$ | |
| 5 | $k_{21} \ll k_{23} \approx k_{12},$ | $1 < \alpha$ | never | N A | |
| 6 | $k_{23} \ll k_{21} \approx k_{12}$ | $\frac{k_2}{k_{12}} < \alpha$ | $(G^*_{23} - G_2) - (G^*_{12} - G_1) \gtrapprox .2 k_B T$ | $\frac{k_{12} k_2}{k_{12} \, k_{21}} \approx \frac{1}{2} k_{23}$ | |

.

The cases 4 and 6 can be analysed similarly. Examples of each of these conditions are given in Table 3.1.

When the conditions are fulfilled we can express the two state kinetics by effective rate constants $k_f$, and $k_b$. The sum of these rate constants is equal to the inverse of the longest timescale, $k_f + k_b = 1/\tau_1 = -\lambda_1$, which in turn corresponds to the smallest eigenvalue. Inserting Eq S32 and approximating for $D \ll 1$ we obtain

$$k_f + k_b = \frac{1}{2}k_{sum}\left(1 - \sqrt{1-4D}\right) \approx k_{sum}D = \frac{k_{12}k_{23} + k_{12}k_{32} + k_{21}k_{32}}{(k_{12} + k_{21} + k_{23} + k_{32})}. \tag{S59}$$

When we, as above, assume that the reverse $k_{32}$ is much smaller than any of the other rates, and by construction also $k_b \ll k_f$, if follows that

$$k_f = \frac{k_{12}k_{23}}{k_{12} + k_{21} + k_{23}}. \tag{S60}$$

This expression, when $k_{12}$ is also small, reduces to the well known steady state expression

$$k_f^{ss} = \frac{k_{12}k_{23}}{k_{21} + k_{23}}. \tag{S61}$$

When $k_{12} \ll k_{23} \approx k_{23}$ we recover case 4, and the effective rate constant is about $k_f = \frac{1}{2}k_{12}$ (see also Table 3.1).

The other case $k_{21} \ll k_{12}, k_{23}$ (case 5) gives

$$k_f = \frac{k_{12}k_{23}}{k_{12} + k_{23}}, \tag{S62}$$

but is in fact never achieved since the 2 state condition is never fulfilled.

The third possibility $k_{23} \ll k_{12}, k_{21}$ leads to

$$k_f = \frac{k_{12}k_{23}}{k_{12} + k_{21}}. \tag{S63}$$

When $k_{23} \ll k_{21} \approx k_{12}$ we recover case 6, and the effective rate constant is about $k_f = \frac{1}{2}k_{23}$ (see also Table 3.1).

Finally, the opposite cases lead to case 1-3. For instance, $k_{12} \gg k_{21}, k_{23}$ yields case 1

$$k_f = \frac{k_{12}k_{23}}{k_{12} + k_{21} + k_{23}} \approx k_{23}, \tag{S64}$$

whereas $k_{23} \gg k_{12}, k_{21}$ yields case 2

$$k_f = \frac{k_{12}k_{23}}{k_{12} + k_{21} + k_{23}} \approx k_{12}, \tag{S65}$$

and finally for $k_{21} \gg k_{12}, k_{23}$ this gives case 3

$$k_f = \frac{k_{12}k_{23}}{k_{12} + k_{21} + k_{23}} \approx \frac{k_{12}.k_{23}}{k_{21}} \tag{S66}$$

These results are also summarised in Table 3.1. We note that the above analysis is indeed in agreement with the results from Kramers' theory.

## 3.2  Time dependent relaxation

Instead of making approximations as was done in the previous section we can also solve the time dependent population numerically, using Eq. S29. We illustrate this for two cases. First, we make the reverse process $k_{32} = 0.000001$ and $k_{21} = 0.000001$, very unlikely (where we take the unit of time dimensionless and arbitrary). The rate across the second barrier is set to $k_{23} = 0.00001$. The $k_{12}$ varies between 0.000001 and 1. We show in figure S7 solutions to Eq. S29, starting from population $p_1 = 1$, and $p_2, p_3 = 0$. Also shown are the corresponding free energy landscapes. The rates are chosen such that for these energy landscapes $G_{12}^* > G_{23}^*$. For each of the settings the intermediate 2 first slowly fills up, followed by a steady (initially linear, later exponential) increase of state 3. In the equilibrium most of the population is in state 3. Note that the population in intermediate 2 reaches steady state fast for $k_{12} < k_{21}$ but overshoots dramatically for the opposite case $k_{12} > k_{21}$, before relaxing back to the steady state value.

Next, we make the reverse process $k_{21} = 0.0001$, less unlikely, while keeping $k_{23} = 0.00001$. The $k_{12}$ varies between 0.000001 and 1. We show in figure S8 solutions to Eq. S29,

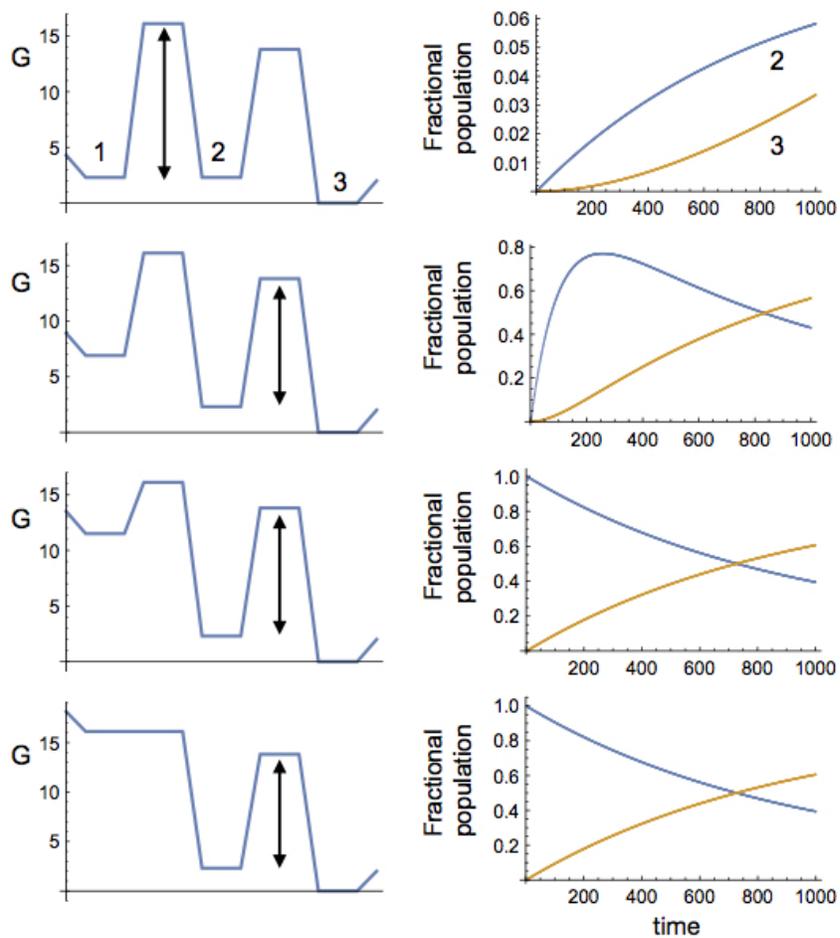

Figure S7: In the right column, the time dependent populations for state 2 (blue) and state 3 (red) are plotted for, from top to bottom, $k_{12} = 0.000001, 0.0001, 0.01, 1$. The unit of time is arbitrary and dimensionless. The left column shows the corresponding free energy landscape. For these free energy landscapes $G_{12}^* > G_{23}^*$

starting from population $p_1 = 1$, and $p_2, p_3 = 0$. Also shown are the corresponding free energy landscapes. The rates are chosen such that for these energy landscapes $G^*_{12} < G^*_{23}$. For each of the settings the intermediate 2 first quickly fills up, followed by a steady (initially linear, later exponential) increase of state 3. In the equilibrium most of the population is in state 3. Note that the population in intermediate 2 reaches steady state fast for $k_{12} < k_{21}$ but overshoots dramatically for the opposite case $k_{12} > k_{21}$, before relaxing back to the steady state value.

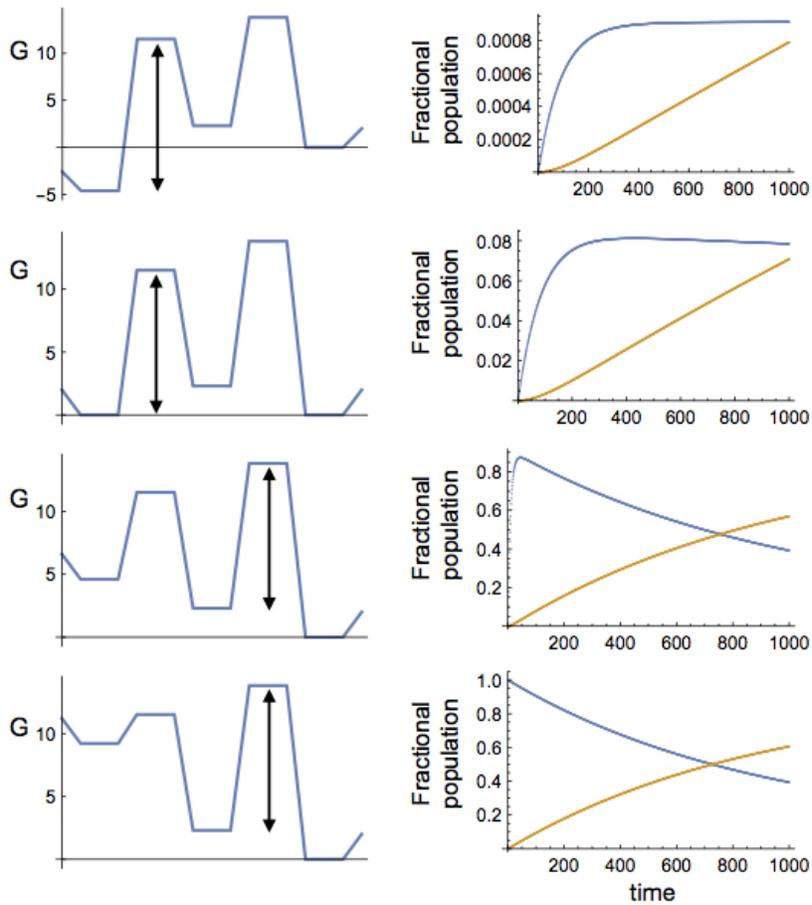

Figure S8: In the right column, the time dependent populations for state 2 (blue) and state 3 (red) are plotted for, from top to bottom, $k_{12} = 0.000001, 0.0001, 0.01, 1$. Unit of time is arbitrary and dimensionless. The left column shows the corresponding free energy landscape. For these free energy landscapes $G^*_{12} < G^*_{23}$



\end{document}